\documentstyle[aps,prb,aps,epsfig]{revtex}
\begin{document}
\draft

\def \rem #1 {{\it [#1] }} 

\def \sm {{s_m^+}}
\def \sk {{s_k^+}}
\def \fk {{F_k}}
\def \fks {{F^2_k}}
\def \fm {{F_m}}
\def \fms {{F^2_m}}
\def \hkm {{G_{km}}}
\def \hkms {{G^2_{km}}}
\def \ha  {{1\over 2}}
\def \l {{\lambda}}
\def \r {{\sqrt{3}}}
\def \neel {{N{\'e}el }}
\def \xxz {{\it XXZ }}
\def \lsubm {{LSUB{\it m} }}

\title{An efficient implementation of high-order coupled-cluster 
techniques applied to quantum magnets} 
\author{Chen Zeng\cite{muetter}, D.J.J. Farnell, and R.F. 
Bishop\cite{muetter2}}
\address{Department of Physics,
University of Manchester Institute of Science and Technology (UMIST), \\
P.O. Box 88, Manchester M60 1QD, United Kingdom}
\date{\today}
\maketitle

\makebox[1cm]{}

\neel model state, square lattice \xxz model, triangular lattice Heisenberg 
antiferromagnet, ket-state parametrisation, bra-state parametrisation, 
lattice animals and fundamental configurations, ground-state energy, 
sublattice magnetisation, anisotropy susceptibility, critical points,
quantum order, quantum phase transitions.

\begin{abstract}
We illustrate how the systematic inclusion of multi-spin correlations 
of the quantum spin-lattice systems 
can be efficiently implemented within the framework of the coupled-cluster 
method by examining the ground-state properties of both the square-lattice
and the frustrated triangular-lattice quantum antiferromagnets. Various 
physical quantities including the ground-state energy, the 
anisotropy susceptibility, and the sublattice magnetisation are 
calculated and compared with those obtained from such other methods 
as series expansions and quantum Monte Carlo simulations. 
\end{abstract}

\section{Introduction} 

The techniques now available in the field of {\em ab initio} 
quantum many-body theory have become increasingly refined over 
the last decade or so. This is particularly true for what is 
nowadays recognised as one of the most powerful modern techniques, 
namely, the coupled-cluster method (CCM)\cite{refc1,refc2,refc3,%
refc4,refc5,refc6,refc7,refc8,refc9}. The results obtained from 
the CCM have become fully competitive with series expansions, 
variational calculations and quantum Monte Carlo (QMC) simulations 
(for the cases in which QMC may be applied).

Quantum magnets not only provide useful models of many physically
realisable magnetic systems but also serve as prototypical  
models of quantum many-body systems. Their rich phase diagrams 
due to strong quantum effects have naturally provided  
an excellent test-bed where the above-mentioned methods can be 
applied and further refined. One example demonstrating rich 
and initially unexpected behaviour is provided by the Haldane
conjecture\cite{Haldane}, which states that the one-dimensional
(1D) spin-1 Heisenberg antiferromagnet (HAF) possesses an 
excitation gap, in sharp contrast to its spin-$\frac 12$ counterpart.
This was surprising at the time because conventional
spin-wave theory predicts a gapless excitation spectrum 
regardless of spin magnitude. However, the Haldane
conjecture has subsequently been confirmed by numerical 
calculations\cite{White}. Moreover, in the aftermath of the 
discovery of the superconducting cuprates, much effort has
been devoted to uncovering such subtle effects as spin-nematic,
spin-Peierls and chiral spin liquid orderings
in two-dimensional (2D) quantum 
antiferromagnets, among which the frustrated quantum antiferromagnets 
on the triangular and the Kagom\'e lattices have recently attracted 
considerable theoretical attention\cite{Kalmeyer,Zeng1,Chalker,%
Singh1,Sachdev,Yang,Bernu,Elstner,Bernu2,Deutscher}.

The CCM has been applied to various quantum magnets over the past
six years. The first application of the CCM to these systems was
performed by Roger and Hetherington\cite{Roger}, who obtained good
results at low levels of approximation for the ground-state energy
of both the 1D chain and the 2D square-lattice HAF, and also for solid
$^3$He where ring exchanges of nuclear spins are considered.
Since then the CCM has been applied to the isotropic (Heisenberg)
and anisotropic HAF (or \xxz model) in 1D and on
the 2D square lattice, both for 
spin-$\frac 12$\cite{Bishop1,Harris,Cornu,Bishop2} systems and 
higher spins systems\cite{Bishop3,Lo1}; to the spin-1 
Heisenberg-biquadratic model\cite{Bishop4}; 
and to such {\em frustrated} spin models as the spin-$\frac 12$
$J_1$-$J_2$ (or Majumdar-Ghosh) model\cite{Farnell1,Xian1,Bursill}
and the 2D triangular lattice HAF\cite{Zeng2,Zeng3}. It has also 
been applied to the spin-1 easy-plane ferromagnet\cite{Lo2}.
Among these, Bishop
{\em et al.}\cite{Bishop1,Bishop4} not only put forth several systematic
{\em localised} approximation schemes to perform higher-order
calculations yielding good results on the ground-state sublattice
magnetisations and approximate excitation spectra, but also used
an infinite-order, two-body approximation scheme to obtain
evidence of a zero-temperature quantum phase transitions.
Moreover, the systematic inclusion of spin-spin correlations based
on a {\em dimerised} state has also been made possible within
the framework of the CCM\cite{Xian1,Xian2}, 
to study spin-Peierls ordering.
This may provide a possible inroad to probe more subtle
topological order in the absence of {\em solid} order, as in the 
case of the 
chiral spin liquid\cite{Kalmeyer,Yang}. The quantitative description
of such phases remains one of the most challenging problems for 
modern microscopic quantum many-body theory in general, and 
the CCM in particular.


More recently, attention has been given to 
extending the CCM calculations to higher orders\cite{Bishop2}
in the particular case of the {\em XXZ} model, by 
using a {\em localised} approximation scheme, and
by taking into account multi-spin correlations on up to $10$
contiguous lattice sites in 1D and on up to
$6$ contiguous lattice sites in 2D. The ground-state energies, 
for example, are found to be in excellent agreement (i.e., within 
about $0.03\%$) with the exact result in 1D, and with those obtained  
from spin-wave theory\cite{Anderson1}, series expansion\cite{Singh2}
and QMC calculations\cite{Carlson,Runge1} in 2D. 
However, it is fair to say that in 2D, to achieve the 
same accuracy on other more interesting physical quantities 
such as the ground-state sublattice magnetisation and 
the excitation spectrum, and to further clarify the nature of  
zero-temperature quantum phase transitions, the inclusion 
of multi-spin correlations of still higher orders is clearly
needed. Since the extent of
the task of determining the CCM equations and solving them 
grows extremely rapidly with the approximation level, efficient 
algorithms for performing the CCM calculations thus become  
indispensable\cite{Zeng3}.   

The motivation of the present work is two-fold: (1) we 
analyse in detail the computational aspects of the CCM         
in the context of quantum antiferromagnets, in order to devise an            
efficient implementation and to show how the systematic 
inclusion of multi-spin correlations can be made simple; 
and (2) we revisit the spin-${1\over 2}$ quantum           
antiferromagnets on both the square and the triangular lattices   
by utilising this algorithm  in a way which now
enables us to increase the number of independent 
and {\em localised} multi-spin configurations to be 
considered by at least an order of magnitude over previous
calculations. In this   
article we focus on the above two models in the regimes          
where a N{\'e}el-like order represents the corresponding    
classical limit. The present method, however, should        
be of general utility to quantum magnets where a spin-%
Peierls order is relevant, for example.
   
A brief description of the contents of this
article now follows. In Sec. II we  
describe how a reformulation of the CCM problem has been achieved. 
The characteristic CCM similarity transform of the spin operators 
is evaluated at a very general approximation level, and the Hamiltonian 
is reformulated purely in terms of spin-raising operators for a 
spin-$\frac 12$ system. We also show that the form of the new 
Hamiltonian easily lends itself to a localised set of approximations 
called the LSUB{\em m} scheme. In so doing the CCM technique is 
itself clarified.
The computational method used to determine our fundamental set of 
configurations within this approximation scheme is described, 
and the derivation of the resulting CCM equations is discussed 
in detail. The method is then applied to the square-lattice 
spin-$\frac 12$ {\em XXZ} antiferromagnet, and the 
triangular-lattice spin-$\frac 12$ anisotropic 
antiferromagnet\cite{Singh1} in Secs. III and IV 
respectively. Our conclusions are given in Sec. V, where we 
also discuss possibilities for further extending our 
computational solution to even higher-order calculations 
by making use of parallel processing and other strategies.

\section{The CCM Formalism for spin-lattice models} 

\subsection{Basic Ingredients of the CCM} 

Since detailed descriptions of the fundamentals of the CCM are available
in the literature\cite{refc1,refc2,refc3,refc4,refc5,refc6,refc7,%
refc8,refc9}, we only highlight the essential ingredients of its 
application here. The exact ket and bra ground-state energy 
eigenvectors, $|\Psi\rangle$ and $\langle\tilde{\Psi}|$, of a 
many-body system described by a Hamiltonian $H$, 
\begin{equation} 
H |\Psi\rangle = E_g |\Psi\rangle
\;; 
\;\;\;  
\langle\tilde{\Psi}| H = E_g \langle\tilde{\Psi}| 
\;, 
\label{eq1} 
\end{equation} 
are parametrised within the single-reference CCM as follows:   
\begin{eqnarray} 
|\Psi\rangle = {\em e}^S |\Phi\rangle \; &;&  
\;\;\; S=\sum_{I \neq 0} s_I C_I^{+}  \nonumber \; , \\ 
\langle\tilde{\Psi}| = \langle\Phi| \tilde{S} {\rm e}^{-S} \; &;& 
\;\;\; \tilde{S} =1 + \sum_{I \neq 0} \tilde{s}_I C_I^{-} \; .  
\label{eq2} 
\end{eqnarray} 
The single model or reference state $|\Phi\rangle$ is required to have the 
property of being a cyclic vector with respect to two well-defined Abelian 
subalgebras of {\it multi-configurational} creation operators $\{C_I^{+}\}$ 
and their Hermitian-adjoint destruction counterparts $\{ C_I^{-} \equiv 
(C_I^{+})^\dagger \}$. Thus, $|\Phi\rangle$ plays the role of a vacuum 
state with respect to a suitable set of (mutually commuting) many-body 
creation operators $\{C_I^{+}\}$, 
\begin{equation} 
C_I^{-} |\Phi\rangle = 0 \;\; , \;\;\; I \neq 0 \; , 
\label{eq3}
\end{equation} 
with $C_0^{-} \equiv 1$, the identity operator. These operators are 
complete in the many-body Hilbert (or Fock) space,  
\begin{equation} 
1=|\Phi\rangle \langle\Phi| + \sum_{I\neq 0} 
C_I^{+}  |\Phi\rangle \langle\Phi| C_I^{-} \; . 
\label{eq4}
\end{equation} 
Also, the {\it correlation operator} $S$ is decomposed entirely in terms 
of these creation operators $\{C_I^{+}\}$, which, when acting on the 
model state ($\{C_I^{+}|\Phi\rangle \}$), create excitations about the model 
state. We note that although the manifest Hermiticity, 
($\langle \tilde{\Psi}|^\dagger = |\Psi\rangle/\langle\Psi|\Psi\rangle$), 
is lost, the intermediate normalisation condition 
$ \langle \tilde{\Psi} | \Psi\rangle
= \langle \Phi | \Psi\rangle 
= \langle \Phi | \Phi \rangle \equiv 1$ is explicitly 
imposed. The {\it correlation coefficients} $\{ s_I, \tilde{s}_I \}$ 
are regarded as being independent variables, even though formally 
we have the relation, 
\begin{equation} 
\langle \Phi| \tilde{S} =
\frac{ \langle\Phi| {\rm e}^{S^{\dagger}} {\rm e}^S } 
     { \langle\Phi| {\rm e}^{S^{\dagger}} {\rm e}^S |\Phi\rangle } \; . 
\label{eq5}
\end{equation} 
The full set $\{ s_I, \tilde{s}_I \}$ thus provides a complete 
description of the ground state. For instance, an arbitrary 
operator $A$ will have a ground-state expectation value given as, 
\begin{equation} 
\bar{A}
\equiv \langle\tilde{\Psi}\vert A \vert\Psi\rangle
=\langle\Phi | \tilde{S} {\rm e}^{-S} A {\rm e}^S | \Phi\rangle
=\bar{A}\left( \{ s_I,\tilde{s}_I \} \right) 
\; .
\label{eq6}
\end{equation} 

We note that the exponentiated form of the ground-state CCM 
parametrisation of Eq. (\ref{eq2}) ensures the correct counting of 
the {\it independent} and excited correlated 
many-body clusters with respect to $|\Phi\rangle$ which are present 
in the exact ground state $|\Psi\rangle$. It also ensures the 
exact incorporation of the Goldstone linked-cluster theorem, 
which itself guarantees the size-extensivity of all relevant 
extensive physical quantities. One crucial difference between 
the CCM parametrisation of the ground state and those used in 
spin-wave\cite{Anderson1} and variational Monte Carlo 
calculations\cite{Huse} is that although they all adopt  
an exponentiated form, the former (CCM) contains spin-raising
operators only.    

The determination of the correlation coefficients $\{ s_I, \tilde{s}_I \}$ 
is achieved by taking appropriate projections onto the ground-state 
Schr\"odinger equations of Eq. (\ref{eq1}). Equivalently, they may be 
determined variationally by requiring the ground-state energy expectation 
functional $\bar{H} ( \{ s_I, \tilde{s}_I\} )$, defined as in Eq. (\ref{eq6}), 
to be stationary with respect to variations in each of the (independent) 
variables of the full set. We thereby easily derive the following coupled 
set of equations, 
\begin{eqnarray} 
\delta{\bar{H}} / \delta{\tilde{s}_I} =0 & \Rightarrow &   
\langle\Phi|C_I^{-} {\rm e}^{-S} H {\rm e}^S|\Phi\rangle = 0 ,  \;\; I \neq 0 
\;\; ; \label{eq7} \\ 
\delta{\bar{H}} / \delta{{s}_I} =0 & \Rightarrow & 
\langle\Phi|\tilde{S} {\rm e}^{-S} [H,C_I^{+}] {\rm e}^S|\Phi\rangle 
= 0 , \;\; I \neq 0 \;\; . \label{eq8}
\end{eqnarray}  
Equation (\ref{eq7}) also shows that the ground-state energy at the stationary 
point has the simple form 
\begin{equation} 
E_g = E_g ( \{s_I\} ) = \langle\Phi| {\rm e}^{-S} H {\rm e}^S|\Phi\rangle
\;\; . 
\label{eq9}
\end{equation}  
It is important to realize that this (bi-)variational formulation 
does {\it not} lead to an upper bound for $E_g$ when the summations for 
$S$ and $\tilde{S}$ in Eq. (\ref{eq2}) are truncated, due to the lack of 
exact Hermiticity when such approximations are made. However, it is clear that 
the important Hellmann-Feynman theorem {\it is} preserved in all 
such approximations. 

We also note that Eq. (\ref{eq7}) represents a coupled set of 
nonlinear polynomial 
equations for the {\it c}-number correlation coefficients $\{ s_I \}$. 
The nested commutator expansion of the similarity-transformed Hamiltonian,  
\begin{equation}  
\hat H \equiv {\rm e}^{-S} H {\rm e}^{S} = H 
+ [H,S] + {1\over2!} [[H,S],S] + \cdots 
\;\; , 
\label{eq10}
\end{equation} 
together with the fact that all of the individual components of 
$S$ in the sum in Eq. (\ref{eq2}) commute with one another, imply 
that each element of $S$ in Eq. (\ref{eq2}) is linked directly to
the Hamiltonian in each of the terms in Eq. (\ref{eq10}). Thus,
each of the coupled equations (\ref{eq7}) is of linked cluster type.
Furthermore, each of these equations is of finite length when expanded, 
since the otherwise infinite series of Eq. (\ref{eq10}) will always 
terminate at a
finite order, provided (as is usually the case) that each term in the 
second-quantised form of the Hamiltonian $H$ contains a finite number of 
single-body destruction operators, defined with respect to the reference 
(vacuum) state $|\Phi\rangle$. Therefore, the CCM parametrisation naturally 
leads to a workable scheme which can be efficiently implemented 
computationally. It is also important to note that at the heart
of the CCM lies a similarity transformation, in contrast with  
the unitary transformation in a standard variational formulation 
in which the bra state $\langle\tilde\Psi|$ is simply taken as
the explicit Hermitian adjoint of $|\Psi\rangle$. 

\subsection{Computational Aspects of the CCM for Spin-Lattice Models} 

\subsubsection{Ket-State CCM Equations} 


In this section, the general formalism of the CCM outlined in the 
previous section will be henceforth applied to spin-$\frac{1}{2}$ quantum  
magnets with the emphasis on the common computational aspects involved 
in its implementation. Further model-specific details are deferred until
Secs. III and IV.

To be specific, in this article we restrict ourselves to spin-${1\over2}$ 
quantum antiferromagnets in the regimes where the corresponding 
classical limit is described by a generalised N\'eel-like ordering, i.e., 
where all spins on each sublattice are separately aligned in the coordinates 
of a global quantisation axis. However, it is a simple task (see Secs. III 
and IV for details) to introduce a different local quantisation axis on 
each sublattice by a suitable spin-rotation transformation, such that 
the above N\'eel-like state becomes a fully aligned (``ferromagnetic'') 
configuration in the local spin coordinates. This ``ferromagnetic'' 
state, $|\Phi\rangle$, will be chosen as the 
uncorrelated CCM model state, where, in the local axes, all
spins point along the respective negative {\em z}-axis,
\begin{equation}
\vert \Phi \rangle =\bigotimes_{i=1}^N \vert
\downarrow \rangle _i\,;
\,\, \text{in the local quantization axes}
\,\, .
\label{eq22-modelstate}
\end{equation}
The correlation operator $S$ is then decomposed
wholly in terms of sums of products of single
spin-raising operators, $s^+_k \equiv s^x_k + i s^y_k$, 
again defined with respect to the local quantisation axes,  
\begin{equation}
S= [i_1] s^+_{i_1} + [i_1 i_2] s^+_{i_1} s^+_{i_2} + \cdot\cdot\cdot
\,\, ,
\label{eq11}
\end{equation}
where $[i_1]$, $[i_1 i_2]$ and so on stand for the corresponding (symmetric)
spin-correlation coefficients (recall $\{ s_I \}$ in Sec. IIA) 
specified by the sets of site indices,  
$\{i_1\}$, $\{i_1, i_2\}$ and so on, on the regular lattices 
under consideration. 
Implicit summations over repeated indices are also
assumed. According to Eq. (\ref{eq7}), the spin-correlation coefficients in
Eq.\ (\ref{eq11}) are to be determined by a set of CCM nonlinear equations:
\begin{equation}
0 = \langle \Phi| s^-_{j_1} s^-_{j_2} \cdot\cdot\cdot s^-_{j_M}
   {\rm e}^{-S} H {\rm e}^{S}
   |\Phi \rangle
\,\, ,
\label{eq12}
\end{equation}
where $s^-_{j_1} s^-_{j_2} \cdot\cdot\cdot s^-_{j_M}$
is the Hermitian conjugate of the corresponding multi-spin
correlation string $s^+_{j_1} s^+_{j_2} \cdot\cdot\cdot s^+_{j_M}$.


In practice we clearly need an approximation scheme to truncate the 
expansion of $S$ in Eq.\ (\ref{eq11}) to some finite or
infinite subset of the full set of multi-spin configurations $\{ I\}$.
The three most commonly used truncation
methods up till now are: (1) the SUB$n$ scheme,
in which all correlations involving only $n$ or fewer spins are retained,
however far separated on the lattice; (2) the simpler SUB$n$-$m$
sub-approximation, where only SUB$n$ correlations spanning a range of no more
than $m$ adjacent lattice sites are retained; and (3) the systematic local
LSUB$m$ scheme, which includes all multi-spin correlations over all
possible distinct locales on the lattice defined by $m$ or fewer contiguous 
sites. Only the last approximation scheme will be adopted throughout this 
article.   

The first step in the practical implementation of the LSUB$m$ CCM is  
to enumerate all of the distinct multi-spin configurations or 
correlated clusters, which we shall 
henceforth call {\em fundamental} configurations, 
$\{i_1, i_2, \cdot\cdot\cdot, i_n \}$ with $n \le m$, retained in the LSUB$m$ 
approximation. It should be noted that the multi-spin configurations that are
related by Hamiltonian symmetries, translational and rotational alike,
are counted as one single distinct configuration. Such a correlated cluster 
can be either a connected cluster of size $m$ 
(also called a ``lattice animal'' 
or ``polyomino'') or a subset of it (connected or disconnected). 
Although the asymptotic behaviour of the number of lattice animals on 
a regular lattice remains an open combinatorial question, efficient 
algorithms for enumerating lattice animals up to sizes of about $20$ 
have been developed in various fields including percolation 
and cell growth problems\cite{Mertens}.

The second step in our modular implementation, namely, 
generating the corresponding set of CCM equations, is what we will 
focus on in the remainder of this section. 
Equation (\ref{eq12}) reveals that there are essentially two
computational aspects involved in obtaining all possible non-zero
contributions to its right-hand side.  The first is to
calculate the similarity-transformed Hamiltonian which then acts
on the model state, and the second is to select terms
of the similarity-transformed Hamiltonian that
match exactly the string of spin-lowering operators represented by
the set of site indices $\{j_1, j_2 ,  \cdots, j_M\}$.
The first aspect is intrinsically related to the noncommutative nature of
quantum spin operators, and the second to the geometric considerations
of the lattice on which the Hamiltonian is defined.
We address these two aspects in more detail below.


The computation of the similarity-transformed Hamiltonian, $\hat{H}
\equiv {\rm e}^{-S} H {\rm e}^{S}$, which acts on the model state 
$|\Phi\rangle$  
can be performed straightforwardly by making use of 
the relations $s^{-}|\Phi\rangle   
= 0$ and $s^{z}|\Phi\rangle=-\frac{1}{2} |\Phi\rangle$.
The goal here is 
to completely eliminate $s^{z}$ and $s^{-}$, and thus retain the 
creation operators only, by utilising the commutation relations 
of the spin operators, namely, $[s^z,s^{\pm}]=\pm s^{\pm}$ 
and $[s^-,s^+]=-2s^z$. 
This greatly simplifies the matching problem in generating 
the CCM equations as discussed below. To this end, we note that 
the similarity-transformed single-spin operators can be expressed as: 
\begin{equation}  
\hat s_k^+  =  \sk \;\; , \;\; 
\hat s_k^z  =  s_k^z + F_k \sk \;\; , \;\;
\hat s_k^-  =  s_k^- -2 F_k s_k^z - (F_k)^2 \sk \;\; ,  
\label{eq13}
\end{equation} 
where $F_k \equiv \sum_l l[k i_1\cdot\cdot\cdot
i_{l-1}] s^+_{i_1}\cdot\cdot\cdot s^+_{i_{l-1}}$.   
Furthermore, the commutation relations between the spin operators 
and the $F_k$ operators can also be written in the following 
compact forms,  
\begin{eqnarray}  
\left\lbrack s_k^z, F_m \right\rbrack
= G_{km} \sk  \;\; &,& \;\;    
\left\lbrack s_k^-, F_m \right\rbrack 
= -2 G_{km} s_k^z \;\; ,  \nonumber \\
\left\lbrack s_k^z, (F_m)^2 \right\rbrack 
= 2 F_m G_{km} \sk \;\; &,& \;\;   
\left\lbrack s_k^-, (F_m)^2 \right\rbrack
= -2 (G_{km})^2 \sk - 4 F_m G_{km} s_k^z \;\; ,  
\label{eq14}
\end{eqnarray} 
where $G_{km} \equiv \sum_l l(l-1)
[km i_1 \cdot\cdot\cdot i_{l-2}]
s^+_{i_1} \cdot\cdot\cdot s^+_{i_{l-2}}$. Unlike in previous 
equations, repeated indices in Eq. (\ref{eq14}) {\em do not} 
imply summations. Clearly both $F$ and $G$ operators contain 
creation operators only.   
Consider then a typical two-spin interaction term, 
$s_k^- s_m^-$, for example, as contained in most spin-lattice Hamiltonians 
(see Sec. III and IV for a full description of the quantum 
spin Hamiltonians actually studied in this article). It is easy  
to prove the following relation:  
\begin{eqnarray} 
\hat{s}_k^-  \hat{s}_m^- |\Phi\rangle  
&=& 
\Bigl(  2(G_{km})^2 \sk\sm + 4 F_k F_m G_{km} \sk\sm 
+ (F_k)^2 (F_m)^2 \sk\sm  \Bigr) |\Phi\rangle \nonumber \\ 
&-& \Bigl( 2 G_{km} F_m \sm + F_k (F_m)^2 \sm 
+ (F_k)^2 F_m \sk +2 G_{km} F_k \sk \Bigr)  
|\Phi\rangle \nonumber \\
&+& \Bigl(  
G_{km}+ F_k F_m \Bigr)  |\Phi\rangle   
\;\; . 
\label{eq15}
\end{eqnarray} 


In Eq. (\ref{eq15}), the resulting terms from $\hat{s}_k^-  \hat{s}_m^-$ are 
classified into three categories as explicitly containing both 
$\sk$ and $\sm$, either $\sk$ or $\sm$, and neither $\sk$ nor $\sm$, 
respectively. The reason for such a classification will become clear 
when we consider the second aspect, namely, generating the CCM equations. 
Thus, the first case is the simplest of all three to deal with,  
since the site indices of all terms in the case, including both $k$ and $m$, 
are completely fixed up to permutations by  the {\it target} set 
$\{ j_1, j_2, \cdot\cdot\cdot, j_M \}$ according to Eq. (\ref{eq12}). 
Although, unlike in the first case, only one of 
$k$ and $m$ in the second case must lie within 
the set $\{ j_1, j_2, \cdot\cdot\cdot, j_M \}$, the search for 
$k$ or $m$ in the matching problem can be easily performed 
once $m$ or $k$ is fixed. This comes about since the two-spin interactions 
with which we mostly deal are usually short-ranged. 
Typical examples are the nearest-neighbour interactions 
where $k$ and $m$ are simply the nearest neighbours as in the two models
considered in this article. By contrast, 
neither index $k$ nor index $m$ in the last case must
belong to the set $\{ j_1, j_2, \cdot\cdot\cdot, j_M \}$. Nonetheless, 
for the LSUB$m$ approximation scheme used here, both $k$ and $m$ must 
lie within a {\it finite} set of indices for which  
$\{ j_1, j_2, \cdot\cdot\cdot, j_M \}$ is a subset.    

To be more specific, let us consider the term in Eq. (\ref{eq15}),  
$F_kF_m$, which can be written explicitly as:
\begin{equation} 
F_k F_m = \sum_{l_1} \sum_{l_2} (l_1+1) (l_2+1) 
[k i_1 \cdot\cdot\cdot i_{l_1}] 
[m n_1 \cdot\cdot\cdot n_{l_2}]       
s^+_{i_1} \cdot\cdot\cdot s^+_{i_{l_1}} 
s^+_{n_1} \cdot\cdot\cdot s^+_{n_{l_2}} 
\label{eq16}
\end{equation} 
where summation over repeated indices is implicitly assumed. 
Therefore, generating the part of the CCM equations due to this particular 
term $F_kF_m$ amounts to determining all possible non-zero contributions 
to $\langle\Phi|s^-_{j_1} \cdot\cdot\cdot s^-_{j_M} F_kF_m|\Phi\rangle$
according to Eq. (\ref{eq16}). This is achieved by partitioning the target 
set $\{ j_1, j_2, \cdot\cdot\cdot, j_M \}$ into two subsets 
$\{ i_1 \cdot\cdot\cdot i_{l_1} \}$ and $\{ n_1 \cdot\cdot\cdot n_{l_2} \}$ 
with $l_1+l_2 = M$, followed by a search for the appropriate $k$ and 
$m$ in a {\em nearby} region that includes 
$\{ i_1 \cdot\cdot\cdot i_{l_1} \}$ and $\{ n_1 \cdot\cdot\cdot n_{l_2} \}$ 
respectively, such that both correlation coefficients 
$[k i_1 \cdot\cdot\cdot i_{l_1}]$ and $[m n_1 \cdot\cdot\cdot n_{l_2}]$ 
are the (symmetry-related) fundamental configurations of a given 
LSUB$m$ approximation. Unlike earlier  
work\cite{Roger,Bishop1,Bishop2} where the maximum 
number of fundamental configurations is limited  
to $100$ or so, the present approach based on partition completely eliminates 
the costly procedure implemented previously for avoiding double occupancies  
of spin-$\frac{1}{2}$ objects, and thus reduces the CPU usage a great 
deal. This optimal implementation becomes possible because all of the terms 
(e.g., $F_kF_m$) in the similarity-transformed Hamiltonian ($\hat{H}$) 
and, more importantly, their explicit structures are completely specified 
by the {\it seemingly tedious} reformulation of $\hat{H}$ in terms of 
$F_k$, $F_m$ and $G_{km}$ operators. 
The CCM ket-state equations so obtained can then be solved by 
the standard Newton-Raphson method.  

\subsubsection{Bra-State CCM Equations} 

According to Eq. (\ref{eq6}) in Sec. IIA, it is necessary to obtain both 
the ket-state correlation coefficients $\{s_I\}$ and the bra-state 
correlation coefficients $\{\tilde s_I\}$ in order to compute a 
general ground-state physical quantity such as the sublattice 
magnetisation. The task of generating the bra-state CCM equations 
(see Eq. (\ref{eq8})) turns out to be simple. Firstly, note that this set 
of coupled equations is linear in $\{ \tilde s_I\}$, as is evident 
in Eq. (\ref{eq8}); secondly, a simple equality,     
${\delta^2 \bar{H}}/{\delta \tilde s_I} {\delta s_I}
={\delta^2 \bar{H}}/{\delta s_I} {\delta \tilde s_I}$, demonstrates that 
the bra-state equations can be readily generated from the already 
obtained CCM ket-state equations by appropriate differentiations. 

Similarly, in the context of spin-${1\over2}$ quantum antiferromagnets,
the $\tilde S$ operator is in general decomposed entirely in
terms of annihilation operators which are 
again defined with respect to local 
quantisation axes:  
\begin{equation} 
\tilde S = 1 + \widetilde {[i_1]} s^-_{i_1} + \widetilde {[i_1 i_2]} 
 s^-_{i_1} s^-_{i_2}+ \cdot\cdot\cdot \;\; , 
\label{eq17}
\end{equation} 
where $\widetilde {[i_1]}, \widetilde {[i_1 i_2]} $ and so on denote 
the corresponding bra-state spin-correlation coefficients (recall 
$\{s_I\}$ in Sec. IIA) specified 
by the sets of site indices $\{i_1\}$, $\{i_1, i_2\}$ and so on, 
which is the analogue of Eq. (\ref{eq11}).  
Moreover, the same LSUB$m$ truncation 
scheme is also adopted. This means that there exists a one-to-one 
mapping between the ket-state and the bra-state fundamental correlation
coefficients which, to ease the burden of notation, are now denoted
as $\{x_r\}$ and $\{\tilde x_r\}$ respectively. Now let 
$\delta \bar{H} /\delta {\tilde x_r}={P}_{r}(x_1,x_2,\cdots,\lambda)=0$ 
denote the $r^{{\rm th}}$ ket-state CCM equation, which is given
in terms of the ket-state 
correlation coefficients and $\lambda$, which stands for other parameters 
included in the Hamiltonian such as anisotropy, for example. Consequently    
$\bar{H}$ may now be written in terms of 
${P}_{r}(x_1,x_2,\cdots,\lambda)$ and the bra-state correlation coefficients,
$\{ \tilde x_{r} \}$ as: 
\begin{equation} 
\bar{H} = P_0(x_1,x_2,\cdots,\lambda) + 
\sum_{r=1}^{N_F} {\tilde x_r} P_r(x_1,x_2,\cdots,\lambda) 
\;\; ,  
\label{eq18}
\end{equation}
where $P_0(x_1,x_2,\cdots,\lambda)$ denotes the zeroth-order CCM term,
i.e., the ground-state energy expression, and $N_F$ is the number of 
fundamental configurations retained 
for a given LSUB$m$ approximation level.  
Therefore, the $s^{{\rm th}}$ bra-state equation may now be
rewritten in terms of $\tilde x_r$ and ${P}_{r}(x_1,x_2,\cdots,\lambda)$:
\begin{equation}
\frac{ \partial P_0(x_1,x_2,\cdots,\lambda)} {\partial x_s} +
 \sum_{r=1}^{N_F} \tilde x_r \frac{\partial {P}_{r}
(x_1,x_2,\cdots,\lambda)  } {\partial x_s} = 0  
\;\; ; \;\; s=1,2,...,N_F.   
\label{eq18-1}
\end{equation}
These linear equations may now be solved by using a standard decomposition
technique for linearly dependent equations, such as the LU decomposition 
method, once the ket-state correlation coefficients $\{x_r\}$ are known. 

Equipped with the above CCM algorithms,
which are readily implemented for 
spin-$\frac{1}{2}$ quantum magnets, we report our new findings on 
ground-state properties of both 
the square- and triangular-lattice spin-$\frac{1}{2}$ quantum 
antiferromagnets in Secs. III and IV, respectively, 
in order to illustrate the 
relative simplicity of the present approach. Similar algorithms have 
also been successfully implemented to study the excitation spectra. 
This will be the subject of a future publication\cite{Farnell2}. 

\section{Spin-$\frac{1}{2}$ \xxz antiferromagnet on the
2D square lattice}
In this section we shall consider the
spin-$\frac 12$ \xxz model on the infinite 
square lattice. The \xxz Hamiltonian is given by,
\begin{equation}
H = \sum_{\langle i,j \rangle} \biggl[ s_i^x s_j^x  +
s_i^y s_{j}^y  + \Delta s_i^z s_{j}^z \biggr]
\; ,
\label{eq:hamiltonian}
\end{equation}
where the sum on ${\langle i,j \rangle}$ runs over all nearest neighbour 
pairs and counts each pair only once. 
The square lattice \xxz model has no exact solution, unlike
its 1D counterpart, although approximate analytical and
numerical calculations have been performed. To put later CCM
calculations in context, we note that the \xxz model
has three regimes: an Ising-like phase characterised by
non-zero \neel order; a planar-like phase in which the spins
in the ground-state wavefunction are believed to lie in
the $xy$ plane; and a ferromagnetic phase. A Monte Carlo
study of the 2D anisotropic Heisenberg model was performed
by Barnes {\em et al.}\cite{Barnes1}. They observed that
the staggered magnetisation in the $z$-direction is non-zero 
for $\Delta>1$, but then appears to become zero below $\Delta=1$. 
They therefore conclude that the critical point is probably
very near to this point. In contrast to this Monte
Carlo calculation, Kubo and Kishi\cite{Kubo1} have used sum rules
to investigate the ground state of this system. They state
that the ground state possesses an off-diagonal long-range
order (LRO) akin to that of the {\em XY}-like state at
small anisotropy, $0.0 < \Delta < 0.13\;$. Also, for
$\Delta > 1.78$ they observe that the system demonstrates
non-zero Ising-like LRO. At $\Delta = -1$ there is a
first-order phase transition to the ferromagnetic phase
for this model.

The isotropic Heisenberg point has been extensively studied
using various approximate methods, and so shall be used
as a test case for the CCM results discussed later in this section.
Runge\cite{Runge1} has performed the most accurate Monte
Carlo simulation to date. This provides a value for 
the ground-state energy per spin of
$-$0.66934(4), and a value for the sublattice magnetisation
which is 61.5$\%$$\pm$0.5$\%$ of the classical value. In
comparison, linear spin-wave theory (LSWT) \cite{Anderson1}
gives a value of $-$0.658 for the ground-state energy, and a 
value for the sublattice magnetisation which is 60.6$\%$ of 
the classical value.

\subsection{The Model State}
We begin the CCM treatment of this spin system by choosing
a suitable model state $|\Phi\rangle$ (for a particular
regime), such that all other possible spin configurations
may be obtained by the application of linear combinations
of products of spin-raising operators to this state. In the
Ising-like regime, characterised by non-zero \neel order, a
natural choice for the model state is the \neel state with
spins lying along the $z$ axis. (For clarity, this state
will be referred to as the $z$-axis \neel model state
throughout this article.) We note however that this model
state is not the best choice for all values of $\Delta$
because the ground-state wavefunction of the \xxz model
in the region $-1<\Delta<1$ is believed to contain only
spins which lie in the $xy$ plane. In this regime, we again
use the classical \neel state, but this time with spins
lying on the {\em x} axis. (This state will be referred
to as the planar model state throughout this article.)
Hence, we see that even for the same spin model and
lattice a different choice of model state may be preferable,
depending on the regime that we are investigating.

We shall consider the Ising-like regime first, and, so
that spins on either sublattice may be treated
equivalently, we peform a rotation of the local axes
of the up-pointing spins by 180$^\circ $ about the
$y$ axis. The transformation is described by,
\begin{equation}
s^x \; \rightarrow \; -s^x, \; s^y \; \rightarrow \;  s^y, \;
s^z \; \rightarrow \; -s^z  \; .
\end{equation}
The model state now appears $mathematically$ to consist
of purely down-pointing spins which is precisely
given by Eq. (\ref{eq22-modelstate}), and the Hamiltonian may
be written in terms of these local axes as,
\begin{equation}
H = -\frac 12 \sum_{\langle i,j \rangle}^N \; \biggl[ \; s_i^+
s_j^+ + s_i^-s_{j }^- + 2 \Delta s_i^z s_{j }^z  \; \biggr] \; .
\label{eq:newH}
\end{equation}
For the planar model state, we again rotate the local
axes of these spins on the separate sublattices such
that all spins appear to be lie along the negative
{\em z} direction which is again given by Eq. (\ref{eq22-modelstate}).
This is achieved by rotating the
axes of the left-pointing spins (i.e., those pointing along
the negative $x$ direction) in the planar model
state by 90$^\circ$ about the {\em y} axis, and by
rotating the axes of the right-pointing spins
(i.e., those pointing along the positive $x$ direction) by
270$^\circ$ about the {\em y} axis. 
(The positive $z$-axis is defined to point directly upwards, 
and the positive $x$-axis is defined to point directly to the right.)
Hence the transformation of the local axes of the left-pointing
spins is described by,
\begin{equation}
s^x \; \rightarrow \; s^z, \; s^y \; \rightarrow \;  s^y,
\; s^z \; \rightarrow \; -s^x  \;.  \label{eq:leftRotation}
\end{equation}
and the transformation of the local axes of the right-pointing
spins is described by,
\begin{equation}
s^x \; \rightarrow \; -s^z, \; s^y \; \rightarrow \;  s^y, \;
s^z \; \rightarrow \; s^x  \; ;  \label{eq:rightRotation}
\end{equation}
The transformed Hamiltonian for the planar model state
is now given by,
\begin{equation}
H = -\frac 14 \sum_{\langle i,j \rangle}^N \; \biggl[ \; (\Delta+1)(s_i^+
s_j^+ + s_i^-s_{j }^-) + (\Delta -1)(s_i^+ s_{j
}^- + s_i^{-} s_{j }^+ )+4 s_i^z s_{j }^z  \; \biggr] \; .
\label{eq:newHamiltonian}
\end{equation}
In the remainder of this section, the
power and flexibility of our new formalism is illustrated
by focussing primarily on the planar model state applied
to the \xxz model on the square lattice. Note however
that equivalent calculations have been undertaken
for the $z$-axis \neel model state, and that a general
explanation is presented here only. A more
detailed explanation of CCM calculations that deal with both 
the ground-state properties and the excitation spectrum using the
new formalism for the \xxz model based on this model
state will be presented in Ref. \cite{Farnell2}.

\subsection{Fundamental Excitation Configurations: Lattice Animals}



\begin{figure}
\epsfxsize=11cm
\centerline{\epsffile{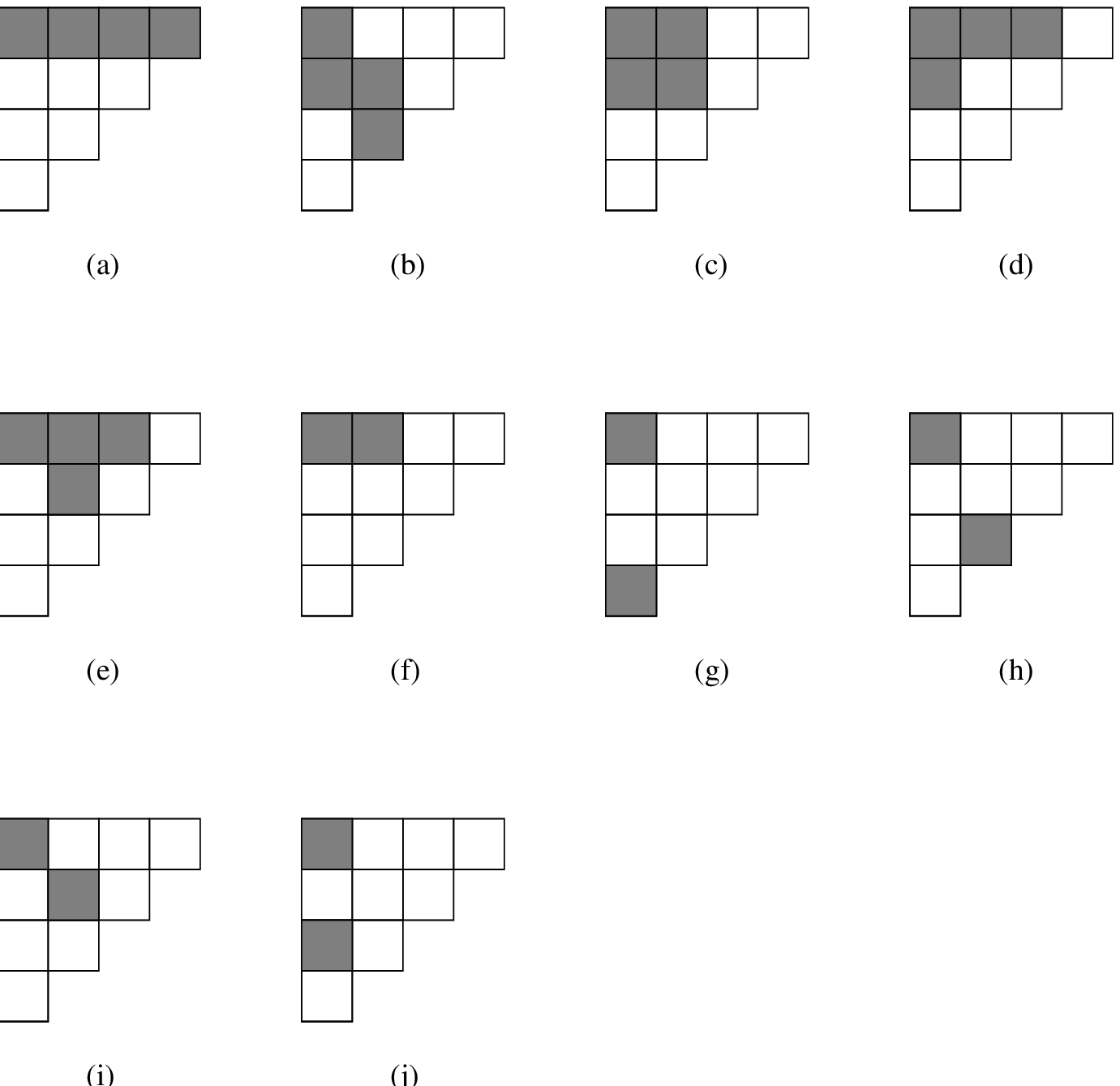}}
\makebox[1cm]{}
\caption{The square lattice LSUB4 bounding triangle is
shown in this figure along with the LSUB4 lattice animals,
illustrated by diagrams (a)-(e). The fundamental LSUB4
configurations for the planar model state are given by
diagrams (a)-(j), and the fundamental configurations
for the $z$-axis \neel model state form a subset of them,
namely, all diagrams except (e), (i), and (j). The centres of 
the shaded squares
mark the relative positions of the sites of the square lattice
on which the spins are flipped with respect to the model
state.}
\label{ba_xxz}
\end{figure}

As described in Sec. IIB,
the first step of our modular solution is to obtain the set of fundamental
configurations for a given approximation scheme by
utilising appropriate lattice symmetries.
Another constructive way to define the LSUB$m$ scheme used here
for the square lattice is to consider a right-angled 
{\it bounding triangle} containing $m$ lattice points 
along the sides parallel to the axes (see Fig.~\ref{ba_xxz} for a
diagram of this construction where the bounding triangle for
LSUB$4$ is shown). All possible fundamental configurations for
the LSUB$m$ approximation are then confined by this bounding 
triangle. This comes
about because it is easy to show that all connected
configurations of size $m$ (or the lattice animals of size $m$)
are constrained to lie within or on this bounding triangle. 
Furthermore, as first
shown by Lunnon\cite{Lunnon}, the introduction of this bounding
triangle greatly simplifies the recursive procedure of
{\it growing} a connected cluster of given size. The disconnected
configurations for LSUB$m$ scheme are then constructed by
successively considering all ``subsets'' of each member of the fundamental
set of connected configurations, and all possible disconnected 
configurations are thereby generated.  (The ``subsets'' here refer to 
all independent configurations which are formed by removing one 
or more spins from these connected configurations.)

To be specific, for the square lattice, there are four
rotational operations, ($0^\circ$, $90^\circ$, 
$180^\circ$, $270^\circ$),
and four reflections, (along the $x$ and $y$ axes,
and along the lines $x$=$y$ and $x$=$-y$), which
preserve the symmetries of both the lattice and the
Hamiltonian. Moreover, the Hamiltonian of Eq.
(\ref{eq:newHamiltonian}), which is defined with respect
to the planar model state, contains only even products
of spin-flip operators and a single term containing two $s^z$ operators.
Repeated application of this Hamiltonian to this model
state yields the ground state (assuming that
this model state is not orthogonal to it). Therefore the
ground state will have an even numbers of spin flips with
respect to this model state, and so we restrict the
\lsubm approximations to contain even numbers of
spin-raising operators in the ket-state correlation operator,
$S$, only for this planar model state. As an example,
in Fig. \ref{ba_xxz} we show all $10$ fundamental
configurations retained in the LSUB$4$ approximation
when the planar model state is used in the CCM calculation.

Further reduction in the number of fundamental configurations
can be made when the $z$-axis \neel model state is used in
the CCM calculations. This comes about because, although
the total uniform magnetisation $s_T^z$=$\sum_i s_i^z$
(where $s_i^z$ is defined with respect to a global
quantisation axis and the sum on the index $i$ runs over all 
lattice sites) is always a good quantum number
independent of the model state used, only the $z$-axis \neel model
state is an eigenstate of the total uniform magnetisation
$s_T^z$. In contrast, the planar model state is not 
an eigenstate of the total uniform magnetisation $s_T^z$.
Therefore, for the $z$-axis model state case one can 
explicitly conserve $s_T^z$
by restricting the fundamental configurations to those
which produce no change in $s_T^z$ with respect to the
$z$-axis \neel model state. This restriction, for example,
reduces the number of the fundamental configurations
retained in the LSUB$4$ approximation to $7$ if
the $z$-axis \neel model state is employed in the CCM calculations,
and see Fig.~\ref{ba_xxz}. We tabulate the number of
of fundamental configurations up to the LSUB$8$ level of
approximation for both model states in Table \ref{tab:2D_xxz}.
Note that only the CCM calculations based on
the $z$-axis \neel model state are actually carried out
up to LSUB$8$ approximation in this article.

\subsection{Similarity-transformed Hamiltonian and CCM
Ket-State Equations}
In order to solve the Schr{\"o}dinger equation of Eq. (\ref{eq1}),
we shall specifically utilise the Hamiltonian of Eq.
(\ref{eq:newHamiltonian}), although a comparable analysis
can also be performed for Eq. (\ref{eq:newH}). The expression
for the ket-state correlation operator of Eq. (\ref{eq11}) is
now used to write the similarity-transformed
Hamiltonian, $\hat H$, of Eq. (\ref{eq10}) in terms of the operators
$F_k$ and $G_{km}$. Furthermore, we subdivide $\hat H$
into three categories as discussed in Sec. IIB
to clarify the problem of finding the CCM equations, where
$\hat H |\Phi \rangle = {\rm e}^{-S} H {\rm e}^S  |\Phi \rangle=(\hat H_1
+ \hat H_2 + \hat H_3 )|\Phi \rangle$, such that:
\begin{eqnarray}
\hat H_1 &=&
\frac 12
\sum_{k\rho}
\left\{
-(\hkm+\fk\fm)
+\frac 14 (\Delta-1)
(\fms+\fks)
\right\}
\sk\sm \;  \nonumber \\
&-&
\frac 18 (\Delta+1)
\sum_{k\rho}
\biggl\{
2\hkms
+4\hkm\fk\fm
+\fks\fms
\biggr\}
\sk\sm
\label{eq:1}  \; ; \\
\hat H_2 &=&
\frac 14
\sum_{k\rho}
\left\{
\fm \sm  + \fk \sk
+ \frac 12 (1 -\Delta)(\fm \sk + \fk \sm)\right\}
\nonumber \\
&+&
\frac 18 (\Delta+1)
\sum_{k\rho}
\biggl\{
(2\hkm+\fk\fm)(\fm\sm+\fk\sk)
\biggr\} \; ;
\label{eq:2} \\
\hat H_3 &=&
-\frac 18 
\sum_{k\rho}
\biggl\{
1+(\Delta+1)
(\hkm+\fk\fm)
\biggr\}
\;\; .
\label{eq:3}
\end{eqnarray}
Note that $k$ runs over all lattice sites and that $m$ is
given by $m\equiv k+\rho$, such that $\rho$ covers all
nearest neighbours to $k$. Hence we see from Eq. (\ref{eq:3})
that the ground-state energy of the \xxz model for the
planar model state is given by,
\begin{equation}
\frac {E_g}{N} = -\frac z8 \biggl[2x_1 (\Delta+1)
+ 1\biggr] \; ,
\label{eq:planarGStateEnergy}
\end{equation}
where $x_1 \equiv [k,k+\rho]$ represents all 
nearest-neighbour, two-body correlation coefficients 
in Eq. (\ref{eq11}) which are equivalent under the
translational and rotational symmetries of the lattice,
and $z$ represents the lattice coordination number
(i.e., the number of nearest neighbours to a
given site).
Note that the ground-state energy of Eq.
(\ref{eq:planarGStateEnergy}) is {\it exact} in the
sense that both the exact expansion for $S$
and any non-trivial approximation of it
will always produce this expression.
\subsection{Results}
\subsubsection{Ground-State Energy}
\begin{figure}
\epsfxsize=11cm
\centerline{\epsffile{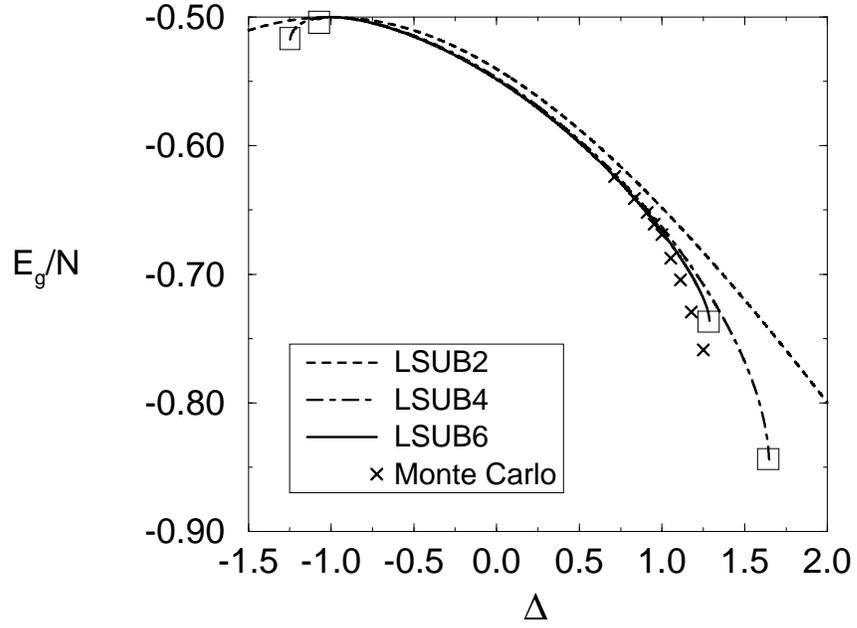}}
\makebox[1cm]{}
\caption{Results for the CCM ground-state energy
of the \xxz model on the 2D square lattice using the
planar model state, compared to the Monte Carlo results of
Ref. [44]. \lsubm critical $\Delta_{c_1}$ and $\Delta_{c_2}$ 
points are indicated by the boxes.}
\label{fig:2D_xxz_energy}
\end{figure}

The ground-state energy for the planar model state is
illustrated by Fig. \ref{fig:2D_xxz_energy}, and we
can see that these results appear to be in good
agreement with Monte Carlo results \cite{Barnes1}.
The highest approximation that has been attempted for
the planar model state is the LSUB6 approximation, which
contains 131 fundamental configurations and gives a
ground-state energy per spin of $-$0.66700 at the Heisenberg
point. For the $z$-axis \neel model state, due to the
reduced number of fundamental configurations,
we were able to solve the LSUB8 approximation,
which contains 1287 fundamental configurations
and gives an energy per spin of $-$0.66817 at $\Delta=1$.
Hence, by utilising the new formalism we have increased
the number of fundamental configurations used
in a CCM calculation for the $z$-axis \neel model state
by over an order of magnitude compared to the previous
best (i.e., LSUB6 \cite{Bishop3} which contains 75
configurations). Table \ref{tab:2D_xxz} summarises
the information regarding numbers of configurations
and ground-state energies at $\Delta=1$.
Note that the calculations for both model
states at the Heisenberg point give exactly the
same results at equivalent approximation levels,
and so only one figure for the ground-state
energy is quoted in Table \ref{tab:2D_xxz}.
The reason for the equivalence is that the
Hamiltonians of Eqs. (\ref{eq:newH}) and
(\ref{eq:newHamiltonian}) become identical at
$\Delta$=1. Also, all of the correlation
coefficients for configurations at a given \lsubm level
contained in the planar model state case but not
contained in the $z$-axis \neel model state case
become identically zero at $\Delta=1$.

The results for the $z$-axis \neel model state are
found to be less accurate than those using the planar
model state in the region $-1<\Delta<1$. Conversely,
for $\Delta>1$, the results based on the $z$-axis
\neel model state become the more accurate of the two
sets of calculations. This therefore vindicates our
decision to use two separate model states in order to
investigate the Ising- and planar-like phases of
this model.
\subsubsection{Anisotropy Susceptibility}
Beyond certain values of the anisotropy parameter (called
critical points) it is found that there is no physically
reasonable solution to the \lsubm CCM equations
for $m\ge4$. This characteristic breakdown of the
solution CCM equations has previously been related
to a phase transition of the real system \cite{Bishop3}.
We may also define a quantity called the anisotropy
susceptibility, given by,
\begin{equation}
\chi_a \equiv - {\partial^2 (E_g/N)\over \partial\Delta^2}
\;\; ,
\label{eq26-1}
\end{equation}
which diverges at the \lsubm critical points. Table
\ref{tab:2D_xxz} illustrates two sets of estimates for the
critical points for the \xxz model based on the planar model
state, corresponding to the ferromagnetic and antiferromagnetic
phase transition points. Encouragingly, the critical
points corresponding to the ferromagnetic phase transition
become closer to $\Delta=-1$ with approximation level. Table
\ref{tab:2D_xxz} also includes the estimates for the critical
points obtained for the $z$-axis \neel model state corresponding
to the antiferromagnetic phase transition point. Note that \lsubm
results based on these two model states always bound the Heisenberg
point, at which the true antiferromagnetic phase
transition is believed to lie, and also appear to converge
with increasing $m$.
\subsubsection{Sublattice Magnetisation}
\begin{figure}
\epsfxsize=11cm
\centerline{\epsffile{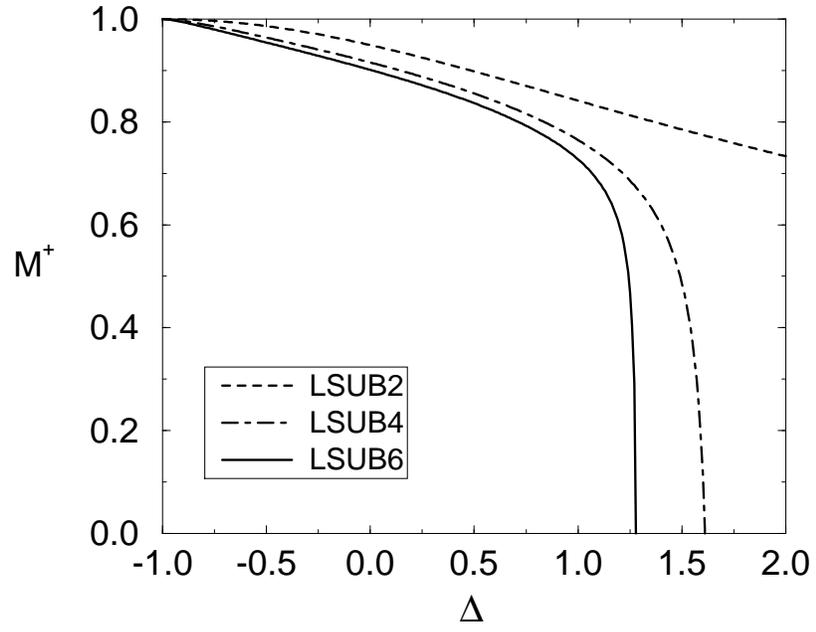}}
\makebox[1cm]{}
\caption{Results for the CCM sublattice magnetisation
of the \xxz model on the 2D square lattice using the
planar model state. CCM results indicate non-zero,
in-plane long-range order in the region $-1<\Delta<1$.}
\label{fig:2D_xxz_mag}
\end{figure}

We now consider a simple order parameter called
the sublattice magnetisation, $M^+\equiv-2\langle s^z\rangle$,
which is defined in terms of the local, rotated spin axes.
Hence $M^+$ is given by,
\begin{equation}
M^+
= \frac {-2}{N_0} \sum_{i=1}^{N_0} \langle\tilde\Psi|
s_i^z|\Psi\rangle
=  1 - \frac 2{N_0} \sum_{k=1}^{N_0} \langle\Phi |\tilde S F_k
\sk|\Phi\rangle
= 1 - 2\sum_{r=1}^{N_F} (n_r)! \: \tilde x_r \: x_r
\;\; ,
\label{eq_mag}
\end{equation}
where the index $i$ runs over all $N_0$
sites of a sublattice, and $N_F$ again
indicates the total number of configurations
for a given \lsubm approximation level. Note that
$\tilde x_r$ and $x_r$, respectively, are 
the bra- and ket-state correlation coefficients
associated with the $r^{{\rm th}}$ fundamental 
configuration, and that $n_r$ is the number
of spins in this configuration.
From Eq. (\ref{eq_mag}) we can see that in order
to obtain a numerical value for the sublattice
magnetisation we must first know the values of
both the ket- and bra-state correlation
coefficients. The manner in which we determine
these coefficients was described in Sec. II.
Our results for the \xxz model using the
planar model state are shown in Fig.
\ref{fig:2D_xxz_mag}, and we note that
these results appear to converge to a non-zero
value as one increases the approximation level.
Hence our results indicate non-zero N{\'e}el-type
long-range order in the $xy$ plane for the
square lattice in the planar regime.
Table \ref{tab:2D_xxz} also summarises the results
for the sublattice magnetisation at the Heisenberg
point. Note that results based on both model
states are again found to be identical at this point
and so only a single number is quoted in Table
\ref{tab:2D_xxz}.

In summary, the new CCM formalism has been used
in order to calculate estimates of the
ground-state energy and sublattice magnetisation
for the 2D \xxz model, and these results
are found to be in good agreement with
other approximate calculations. The
highest-order approximation for the $z$-axis
\neel model state has been extended
to LSUB8 level using the new formalism, which
is an increase of over an order of magnitude
in the number of fundamental configurations
used in the previous-highest LSUB6 calculation.
Also, the positions of the phase transition points
obtained by the CCM, using both model states,
are fully consistent with the known behaviour for
this model. Finally, the results presented here
also support the idea that this model contains
both Ising- and planar-like phases.

\section{Spin-$\frac{1}{2}$ Triangular Lattice Antiferromagnet}

Unlike the square-lattice spin-$\frac{1}{2}$ HAF 
where various calculations including 
extensive quantum QMC simulations\cite{Carlson,Runge1} 
strongly support the existence of a N\'eel ordering  
with a reduced magnetic moment of about $62\%$ 
of its classical value (See Sec. III), the three-sublattice 
ordering for the corresponding triangular 
case is much less clear. For instance, early variational wavefunction 
calculations\cite{Huse} that include long-ranged two-spin and 
nearest-neighbour three-spin correlations support an ordered ground
state with a value of sublattice magnetisation, $M^+=0.68$, i.e., 
as large as $68\%$ of the classical value. Based on this antiferromagnetic 
correlated trial wavefunction, fixed-node Green function 
Monte Carlo simulations were recently performed on lattices 
of up to $324$ sites\cite{Boninsegni}. These  
yielded a similar magnetisation, $M^+ \approx 0.60$\cite{Boninsegni}. 
However, series expansion calculations\cite{Singh1}, utilising 
up to $11$th-order terms in an Ising-like anisotropy parameter 
suggest that the triangular HAF may be at, 
or at least close to (with the magnetisation being extrapolated to 
a value of $M^+ \approx 0.20$), the critical point of losing magnetic 
order. This scenario has received support from exact diagonalisation 
calculations on lattices of up to $36$ sites\cite{Leung}. Yet another 
careful analysis \cite{Bernu2} of the same data from exact diagonalisations  
in terms of a consistent description of the symmetries and 
dynamics of the quasi-degenerate joint states indicates the 
presence of sublattice magnetic order with the 
magnetisation value, $M^+ \approx 0.50$, which is also consistent 
with second-order spin-wave calculations\cite{Deutscher}. 
Clearly, further work is still needed to account for the 
discrepancy, and to provide a more definite and converged result. 

Hence, in this section we therefore further apply the CCM to 
the spin-${1\over 2}$ triangular HAF and focus on model-specific
details of the algorithm implementation discussed in Sec. II.  
Compared with earlier applications of the CCM on the triangular
HAF which only include two-spin (though long-ranged)
correlations\cite{Zeng2}, the current calculations, which 
take into account all multi-spin correlations on up to six 
contiguous sites, have obtained various ground-state properties    
that are now found to be fully competitive with those obtained 
from the above-mentioned methods.

\subsection{Model Hamiltonian and Model State} 

The spin-$\frac{1}{2}$ triangular HAF is described by the
antiferromagnetic-coupling Hamiltonian, 
\begin{equation} 
H=\sum_{\langle i,j\rangle} \vec s_i \cdot \vec s_j 
\;\; ,  
\label{eq19}
\end{equation} 
where $\vec s_i$ denotes the spin-$\frac{1}{2}$ operator at site $i$
on the infinite triangular lattice.  
The sum in Eq. (\ref{eq19}) on $\langle i,j\rangle$ runs over 
all nearest-neighbour pairs and counts each pair once. 
We note that the operators in Eq. (\ref{eq19}) are defined in terms of 
some global 
spin quantisation axes referring to all spins, whereas henceforth 
we shall consistently employ a notation in which the spin operators 
are described in terms of local (N\'eel-like) quantisation axes
for each of the three sublattices (A, B, and C) of the triangular 
lattice. The classical ground-state of Eq. (\ref{eq19}) is the N\'eel-like 
state where all spins on each sublattice are separately aligned (all 
in the $xz$-plane, say). The spins on sublattice A 
are oriented along the negative {\em z}-axis, and spins on sublattices 
B and C are oriented at $+120^\circ$ and $-120^\circ$, respectively, 
with respect to the spins on sublattice A. In order both to facilitate 
the extension of the isotropic Heisenberg antiferromagnet to include 
an Ising-like anisotropy first introduced by Singh and Huse\cite{Singh1}  
and to make a suitable choice of the CCM model state, we perform 
the following spin-rotation transformations. 
Specifically, we leave the spin axes on sublattice A unchanged, and we 
rotate about the $y$-axis the spin axes on sublattices B and C by 
$-120^\circ$ and $+120^\circ$ respectively, 
\begin{eqnarray} 
s_B^x \rightarrow -\frac{1}{2} s_B^x - \frac{\sqrt{3}}{2} s_B^z \;\;  &;& \;\; 
s_C^x \rightarrow -\frac{1}{2} s_C^x + \frac{\sqrt{3}}{2} s_C^z \;\; , 
\nonumber \\
s_B^y \rightarrow s_B^y \;\; &;& \;\; s_C^y \rightarrow s_C^y \;\; , 
\nonumber \\ 
s_B^z \rightarrow  \frac{\sqrt{3}}{2} s_B^x -\frac{1}{2} s_B^z \;\; &;& \;\; 
s_C^z \rightarrow -\frac{\sqrt{3}}{2} s_C^x -\frac{1}{2} s_C^z \;\; . 
\label{eq20}
\end{eqnarray} 
We may rewrite Eq. (\ref{eq19}) in terms of spins defined in these 
local quantisation 
axes for the triangular lattice with a further introduction of an 
anisotropy parameter $\lambda$ for the non-Ising-like pieces, 
\begin{eqnarray} 
H = \sum_{\langle i\rightarrow j\rangle}
\Bigl\{ 
&& 
-{1\over 2} s_i^z s_j^z
+\frac{\sqrt{3}\lambda}{4}
\left( s_i^z s_j^+ +s_i^z s_j^- -s_i^+ 
s_j^z- s_i^-s_j^z \right) \nonumber \\  
&& 
+\frac{\lambda}{8}
\left( s_i^+s_j^- + s_i^- s_j^+ \right) 
-\frac{3\lambda}{8}
\left( s_i^+ s_j^+ + s_i^- s_j^- \right) 
\Bigl\}  
\;\; ,   
\label{eq21}
\end{eqnarray}
where $\lambda=1$ corresponds to the isotropic Heisenberg Hamiltonian 
of Eq. (\ref{eq19}). We note that the summation in Eq. (\ref{eq21}) 
again runs over nearest-neighbour
bonds, but now also with a {\it directionality} indicated by 
$\langle i \rightarrow j\rangle$, which goes from A to B, B to C, and 
C to A. When $\lambda=0$, the Hamiltonian in Eq. (\ref{eq21}) describes 
the usual 
classical Ising system with a unique ground-state which is simply 
the fully aligned (``ferromagnetic'') configuration in the local 
spin coordinates described above. We choose this state as the 
uncorrelated CCM model state $|\Phi\rangle$ which is, of course, 
precisely given by Eq. (\ref{eq22-modelstate}).   

\subsection{Fundamental Excitation Configurations: Lattice Animals} 

\begin{figure}
\epsfxsize=11cm
\centerline{\epsffile{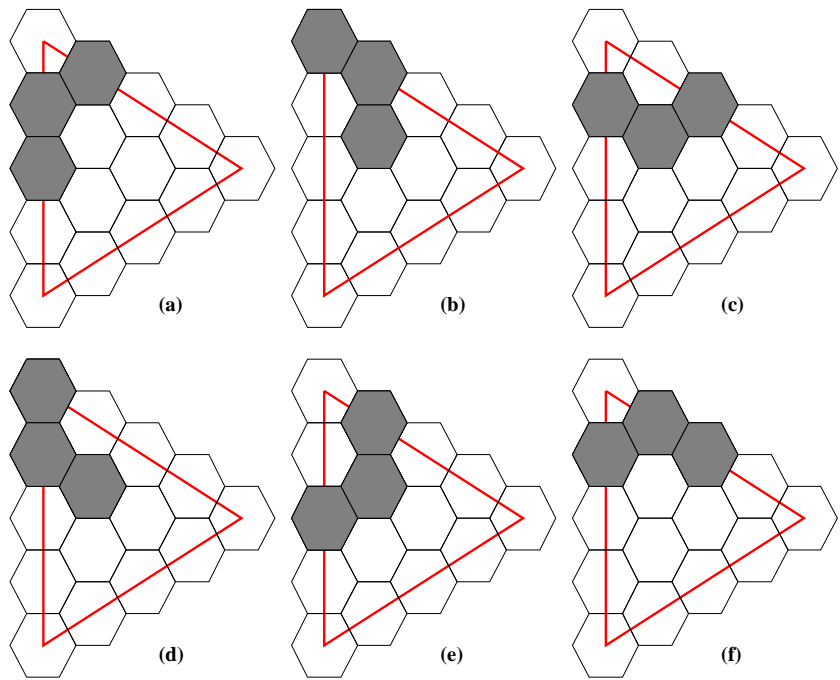}}
\makebox[1cm]{}
\caption{The LSUB$5$ bounding triangle and symmetry-related correlation 
configurations for the triangular lattice.
The centres of the shaded hexagons mark the relative position of the sites of
the triangular lattice on which the spins are flipped with
respect to the model state. See text for details.}
\label{fig4}
\end{figure}

Unlike the square lattice case discussed in Sec. IIIB where all lattice 
point-group symmetries are employed to produce 
symmetry-distinct configurations, care must be exercised here since 
not all of the lattice point-group symmetries leave the 
lattice-spin Hamiltonian invariant. The Hamiltonian of 
Eq. (\ref{eq21}) (or the CCM model state) explicitly breaks some of the 
lattice symmetries because of the presence of  
bond-directionality in the Hamiltonian. 
Thus only $6$ (instead of the full $12$) point-group symmetries 
should be used in the symmetry reduction. These are, specifically, 
three rotational operations ($0^\circ, 120^\circ$, and $240^\circ$) 
together with three reflections about the lattice axes 
(i.e., lines that coincide with the edges of the triangular
lattice). 
For example, the three configurations (a), (b), and (c) shown in 
Fig.~\ref{fig4} are symmetry equivalent, as are the three configurations 
(d), (e), and (f). However, the former are regarded as inequivalent
to the latter in the context of the present spin-lattice Hamiltonian 
problem. 
For the purpose of comparison with the case in, say, percolation
problems, and for the sake of concreteness, let us consider the connected
configurations of size $6$. If all $12$ point-group symmetries were used,
we would have obtained $82$ symmetry-inequivalent configurations as shown
in Fig. 5, which are further classified into two groups: $17$ in group
group A and $65$ in group B. The configurations in group A are of
higher symmetries than those in group B, and thus do not lead to new
symmetry-inequivalent configurations when only $6$ point-group
symmetries must be used in the symmetry reduction as discussed above.
Each configuration in group B, however, results in another new
symmetry-inequivalent configuration. Therefore, the total number
of symmetry-inequivalent connected configurations of size $6$
is $147$ for the present spin-Hamiltonian problem. The set of
symmetry-inequivalent configurations (connected and disconnected)
thereby forms the set of fundamental configurations.
We tabulate the number of fundamental configurations up to the LSUB$7$ 
level of approximation in Table \ref{t1}.  
Note that only CCM computations up to the LSUB$6$ level 
of approximation are actually carried out in this article.   
   
\subsection{Similarity-Transformed Hamiltonian and CCM Ket-State Equations} 

Using Eq. (\ref{eq14}) given in Sec. IIB, we can straightforwardly carry 
out the similarity transformation of the Hamiltonian given by 
Eq. (\ref{eq21}); the resulting terms are further classified into 
three categories for reasons already discussed in Sec. IIB, 
i.e., $\hat H |\Phi\rangle  \equiv
{\rm e}^{-S} H {\rm e}^S |\Phi\rangle
= (\hat H_1 + \hat H_2 + \hat H_3)|\Phi\rangle$,
as indicated below:
\begin{eqnarray} 
\hat H_1 &=& {1\over 4} 
\sum_{k\rho}
\left\{
-2(\hkm+\fk\fm)
-{3\l\over 2}
+{\r\l}(\fk-\fm)(1+2\hkm+\fk\fm)
\right\} \sk\sm \nonumber \\
&+& {1\over 4}
\sum_{k\rho}
\left\{
-{\l\over 2}(\fms+\fks)
-{3\l} \hkms
-6{\l}\hkm\fk\fm
-{3\l\over 2}\fks\fms
\right\} \sk\sm  \; ;
\label{eq23}  \\
\hat H_2 &=& {1\over 4}
\sum_{k\rho}
\left\{
[\fm
-{\r\l\over 2}(1-\fms)
+{\l\over 2}\fk] \sm
+[\fk
+{\r\l\over 2}(1-\fks)
+{\l\over 2}\fm]\sk
\right\}  \nonumber \\
&+& {1\over 4}
\sum_{k\rho}
\left\{
{\r\l}(\hkm+\fk\fm)(\sk-\sm)
+{3\l\over 2}(2\hkm+\fk\fm)(\fm\sm+\fk\sk)
\right\} \; ;
\label{eq24} \\ 
\hat H_3 &=& {1\over 4}
\sum_{k\rho}
\left\{
-\ha
-{\r\l\over 2}(\fm-\fk)
-{3\l\over 2}(\hkm+\fk\fm)
\right\} 
\; .
\label{eq25}
\end{eqnarray}
Here the summation over $k$ runs over all triangular 
lattice sites, while the summation over $\rho$ is over 
the three directed nearest-neighbour vectors that point from A to B, 
B to C, and C to A as required by the explicit bond
directionality in the Hamiltonian given by Eq. (\ref{eq21}),
and the index $m\equiv k+\rho$. 
Each fundamental configuration for a given LSUB$m$ approximation 
is then {\it pattern-matched} to each of the total $35$ terms 
in the above similarity-transformed Hamiltonian following  
the procedure outlined in Sec. IIB to generate the entire 
set of coupled CCM ket-state equations. The 
CCM ket-state equations may then be solved by the Newton-Raphson 
method for nonlinear equations. Specifically, we start from 
the point $\lambda=0$, at which we know the exact solution 
where all the correlation coefficients are zero. We then use 
this known solution as an initial input in solving the CCM 
equations for a slightly increased nonzero anisotropy $\lambda$.
This procedure is carried out recursively to obtain the 
numerical results reported below.  

\subsection{Results}

\subsubsection {Ground-State Energy}
In Fig.~\ref{fig6} we show the ground-state energy per spin $E_g/N$ as  
a function of the anisotropy parameter $\lambda$ for various 
LSUB$m$ approximations. The corresponding values at the isotropic 
Heisenberg point are also tabulated in Table \ref{t1}. The highest-order 
calculation, LSUB$6$, which consists of $758$ independent fundamental 
correlation coefficients, yields $E_g/N=-0.54290$. This value
should be compared with the value $-0.5445$ extrapolated from 
finite-cluster diagonalisations of up to 36-spin clusters\cite{Bernu}, 
and the value $-0.5431\pm0.0001$ from a 
recent QMC simulation\cite{Boninsegni}. 
Compared with the corresponding classical 
value of $-0.375$, it is safe to say that the LSUB$6$ CCM 
calculation captures at least $99\%$ of the quantum corrections. 
Due to the lack of rigourous finite-size scaling results, 
the investigation of which itself 
presents an interesting and important open question, no attempt at 
extrapolation is made here. Nonetheless, the ground-state energy 
obtained from the LSUB$6$ approximation alone is fully competitive 
with those obtained from the above-mentioned alternative methods, 
and is clearly among the best estimates currently available. 

To make further contact with the highest-order series 
expansion known to date\cite{Singh1}, we have computed the 
perturbative solution 
of $E_g/N$ in terms of the anisotropy parameter $\lambda$.  
In Table \ref{t2} we tabulate the expansion coefficients from the
LSUB$6$ approximation, together 
with the corresponding results from exact series expansions\cite{Singh1}.
We note that the LSUB$6$ approximation reproduces the exact series 
expansion up to 
the $6$th order. This result lends further strong support to the 
conjecture that the LSUB$m$ approximation reproduces the {\it exact} series 
expansion to the same $m$th order\cite{Bishop5}. Moreover, the fact that 
the corresponding values of several of the higher-order expansion 
coefficients from both the CCM LSUB$6$ perturbative solution 
and the exact series expansion remain close to each other
shows that the exponential 
parametrisation of the CCM with the inclusion of multi-spin 
correlation up to certain order also captures the dominant 
contributions to correlations of a few higher orders in the 
series expansions. Further detailed series analysis, using such 
methods as 
Pad{\'e} approximant techniques, for the perturbative expansions of both 
the ground-state energy $E_g/N$ and the sublattice 
magnetisation $M^+$ will be reported elsewhere\cite{Zeng4}. 

\subsubsection {Anisotropy Susceptibility} 

As already displayed in Fig.~\ref{fig6}, the LSUB$m$ 
ground-state energy curve for $m \ge 3$ terminates at lower and upper
critical values of the anisotropy, $\lambda_{c_1}$ and $\lambda_{c_2}$
respectively, beyond which no physical solution of the CCM ket-state 
equations exists. We have also tabulated $\lambda_{c_1}$ and $\lambda_{c_2}$
for various LSUB$m$ approximations in Table \ref{t1}. Although the 
CCM based on the model state given in Eq. (\ref{eq22-modelstate}) 
with the three-sublattice magnetic ordering is bound to break 
down in the region of the anisotropy parameter space where the true  
ground-state wavefunction possesses a different symmetry from that 
of the model state, it has been strongly argued\cite{Bishop1,Bishop2} that 
the terminating points may correspond to the critical points of 
a phase transition.  In order to shed further light on the nature of 
the terminating points, we investigate the singular behaviour of the 
CCM correlation coefficients near these points by calculating the 
derivatives of these coefficients with respect to the anisotropy 
parameter. Specifically, we display in Fig.~\ref{fig7} the 
anisotropy susceptibility defined analogously to Eq. (\ref{eq26-1})
as, $\chi_a \equiv - {\partial}^2 (E_g/N)/ \partial {\lambda}^2$.  
Clearly, $\chi_a$ diverges at the corresponding terminating 
anisotropy parameters $\lambda_{c_1}$ and $\lambda_{c_2}$. 
Note that the lower terminating point $\lambda_{c_1}$ clearly 
converges to a value of about $-0.5$, as argued by Singh and 
Huse\cite{Singh1}. However, the upper terminating 
point $\lambda_{c_2}$, as tabulated in Table \ref{t1},  
remains considerably larger than the value of unity which was obtained via 
Pad\'e analysis of the series expansion by Singh and Huse\cite{Singh1}. 
Further work is needed to explain this disagreement and also to 
determine the singularity exponents of the ground-state energy 
at the two critical points.
 
\subsubsection {Sublattice Magnetisation}

Once the ket- and bra-state correlation coefficients 
are known it is possible to evaluate
the sublattice magnetisation, $M^+ \equiv -2 \langle s^z \rangle$,
which is similarly defined by Eq. (\ref{eq_mag}) except that 
the subscript {\em i} now covers all $N_0$ sites on the  
sublattice A, one of the three sublattices. 
In Table \ref{t1} we also tabulate the sublattice magnetisations 
at the Heisenberg point for the various LSUB$m$ approximations. 
The highest-order LSUB$6$ approximation gives give rise to 
a value of $0.6561$, which agrees well 
with the corresponding value of $0.60$ obtained recently from 
QMC simulations\cite{Boninsegni}. However, the discrepancy 
with that from finite-size and the second-order spin-wave 
calculations still needs to be accounted for.    

The divergence in $\langle s^z \rangle$ seen in Fig.~\ref{fig8} 
near the critical points is a natural consequence of the approximate
nature of the calculation. As we approach a critical point for a given 
LSUB$m$ approximation, one of the CCM correlation coefficients $x_r$ 
becomes very large. The contribution to $\langle s^z\rangle $ from this 
coefficient also becomes very large and so the sublattice
magnetisation diverges. The puzzling ``upturn'' of $M^+$ observed 
for the LSUB$5$ and the LSUB$6$ approximations near their
respective upper critical points $\lambda_{c_2}$ remains elusive 
to us at the present. 

In summary, compared with earlier applications of the CCM on the
triangular HAF which have already revealed interesting 
oscillatory behaviour in long-ranged two-spin correlations\cite{Zeng2}, 
the present high-order CCM calculations (with a systematic inclusion 
of multi-spin correlations on up to six contiguous lattice 
sites) have already obtained results including ground-state 
energy and sublattice magnetisation that are fully competitive 
with those obtained from other methods, and which are among the best 
available. Further detailed analysis of the large number of 
ket-state coefficients already obtained for the LSUB6 approximation,  
in order to achieve a better understanding of the nodal surface 
of the ground-state wavefunction, may provide a more microscopic 
justification of, and an extension to, the above-mentioned 
variational wavefunction. This may in turn lead to a better 
trial wavefunction for QMC simulations.

\section{Conclusions and Outlook}

In conclusion to this article, we restate the main points of the new 
formalism, and discuss the results for the two spin models to which 
it has been applied. We also consider the future development of the CCM 
using the new approach, and make suggestions regarding possible 
strategies in order to extend the CCM calculations to even higher orders.

Our new formalism has reformulated the Hamiltonian purely in terms of 
spin-raising operators acting on some suitable model state. We note that 
the reduction 
in CPU processing time is realised by cutting out one large task, namely, 
the evaluation of the similarity transform. Also, the method presented 
here largely negates the task of checking for double occupancy for the 
spin-$\frac 12$ system, which can also cause considerable delay. The 
simple and straightforward nature of the terms within the 
similarity-transformed Hamiltonian means that one may easily amend the 
code in our programs to deal with other systems and lattices, and so we 
have a very flexible approach. 

We have seen that the ground-state energy results for both of the 
systems considered are fully competitive with QMC energies, without 
recourse to large computing facilities; we have used a HP series 
700 workstation containing 96 Megabytes of RAM to perform the 
calculations presented here. We have also seen that sublattice 
magnetisations have been determined which are again fully competitive 
with QMC calculations. 

The present limitations of the new method arise form two sources. One 
limitation
is that we must save our CCM equations in memory or on disk in order 
to obtain expectation values. The rapid increase in the number of 
fundamental configurations with increasing LSUB{\em m} approximation 
level means 
that the cost in memory and disk space rapidly becomes prohibitive, and the 
task of solving the CCM equations becomes slower also. Another 
limitation is that, as discussed in Sec. IIB, we search over a 
wide area in order to determine the `free' indices within specific 
terms in $\hat H$. In 1D, we find that this area depends on $m$ linearly, 
but in 2D, for example, this area increases quadratically with $m$. 
We need to perform this search because, as yet, we do 
not take into account the manner in which the members of the set 
of fundamental configurations are inter-related. For example, 
some configurations are related to others by the addition or
subtraction of a single spin. Thus, most of the processing time 
is spent sweeping through this area regardless 
of whether we know that a particular configuration will contribute
or not. A useful gain could be to determine the way in which the
configurations are related, using graph theory\cite{Flaming} for example,
and then to use this knowledge in order to reduce the search area.

We have seen that the CCM LSUB$6$ approximation fully reproduces 
the first $6$ coefficients of the exact series expansion for the 
triangular HAF. Also, for higher orders in the series than sixth 
order, it is found that the CCM captures about $99\%$ of the 
expansion coefficient of the $7$th order, and about $75\%$  
of those of higher orders up to $11$th order in the   
ground-state energy perturbation 
series on the triangular lattice. CCM expansions of the 
ground-state energy have also been performed up to sixtieth 
order, at a fraction of the computational cost compared to a 
direct calculation of the exact series to this order. A possible 
way of overcoming the memory and disk space limit, when finding 
the CCM equations, is to derive a perturbative solution to the 
CCM equations (as previously discussed in Sec. IV). We obtain 
the expansion coefficients of the ket-state correlation 
coefficients in terms of powers of some coupling parameter. As 
shown in Table \ref{t2}, we may then substitute the series 
expansions for the ket-state correlation coefficients into the 
ground-state energy equation, thus forming a power series for 
this quantity in powers of the coupling parameter. The prospective 
gain is that we might use less disk space in evaluating and 
storing the series coefficients than in storing the CCM equations. 
However, each CCM equation must be evaluated many times in order 
to determine the power series, which in turn means that we must 
significantly increase the speed of our code. We note that the 
number of \lsubm configurations that is needed in order 
to reproduce exact $n^{{\rm th}}$-order perturbation theory is 
a small fraction of the total number of \lsubm configurations.
Hence, if one can state beforehand which configurations
are necessary in order to reproduce $n^{{\rm th}}$-order perturbation 
theory, an elegant method of reducing the processing time and 
memory overheads might be to use this small fraction instead of 
the total number of \lsubm configurations.


Another exciting prospect in order to increase the speed might be to generate 
the CCM equations {\em in parallel}. Also, the CCM is well suited to 
parallelisation as each CCM equation could be implemented on a separate 
processor. Note that the evaluation of each equation depends only on
the target configuration and the form of the Hamiltonian, which 
remains constant. The possible weaknesses of this approach are that one 
is limited not only by the number of nodes in the system, but more 
importantly also by the ``weakest link''. By this we mean that one can only 
progress as fast as the evaluation of the slowest equation at each 
iteration. However, very large gains are possible by sharing out 
the processing needs.


To recap, we have mentioned that our results using the new CCM 
formalism are now fully competitive with QMC results, and we 
have seen that our approach is easily generalisable to other systems. 
Possible future systems to which we may apply the new formalism 
include the valence-bond solids\cite{Xian1,Xian2}, systems with 
higher spin\cite{Lo1,Bishop4}, and models with electronic degrees of 
freedom such as the Hubbard model\cite{Roger2}. The nature of the 
ground states 
of new and interesting materials or spin models might be quickly and 
easily investigated by specifying the Hamiltonian, lattice and spin 
number, and development of the code might therefore lead to a powerful 
test-bench for various new ideas. The future also holds the possibility 
of very high-order calculations, which will increase our knowledge 
both of these systems and also of the CCM. With the inclusion 
of very high orders of approximation it is also hoped that the asymptotic 
nature of the CCM ground-state energies, sublattice magnetisations 
and phase transitions will become clear. In conclusion, we believe that 
the application of the CCM to the lattice spin systems is yielding 
excellent and interesting results, though there are still many 
avenues of fruitful and challenging research to be investigated.
   
\section{Acknowledgements}%

We thank J.B. Parkinson and N.R. Walet for their interesting and useful 
discussions. C. Zeng gratefully acknowledges support from NSF grant 
DMR-9419257 at Syracuse University, and R.F. Bishop gratefully acknowledges 
a grant from the Engineering and Physical Sciences Research Council (EPSRC) 
of Great Britain.



\pagebreak

\newpage

\newpage

\newpage

\newpage
\begin{figure}
\epsfxsize=11cm
\centerline{\epsffile{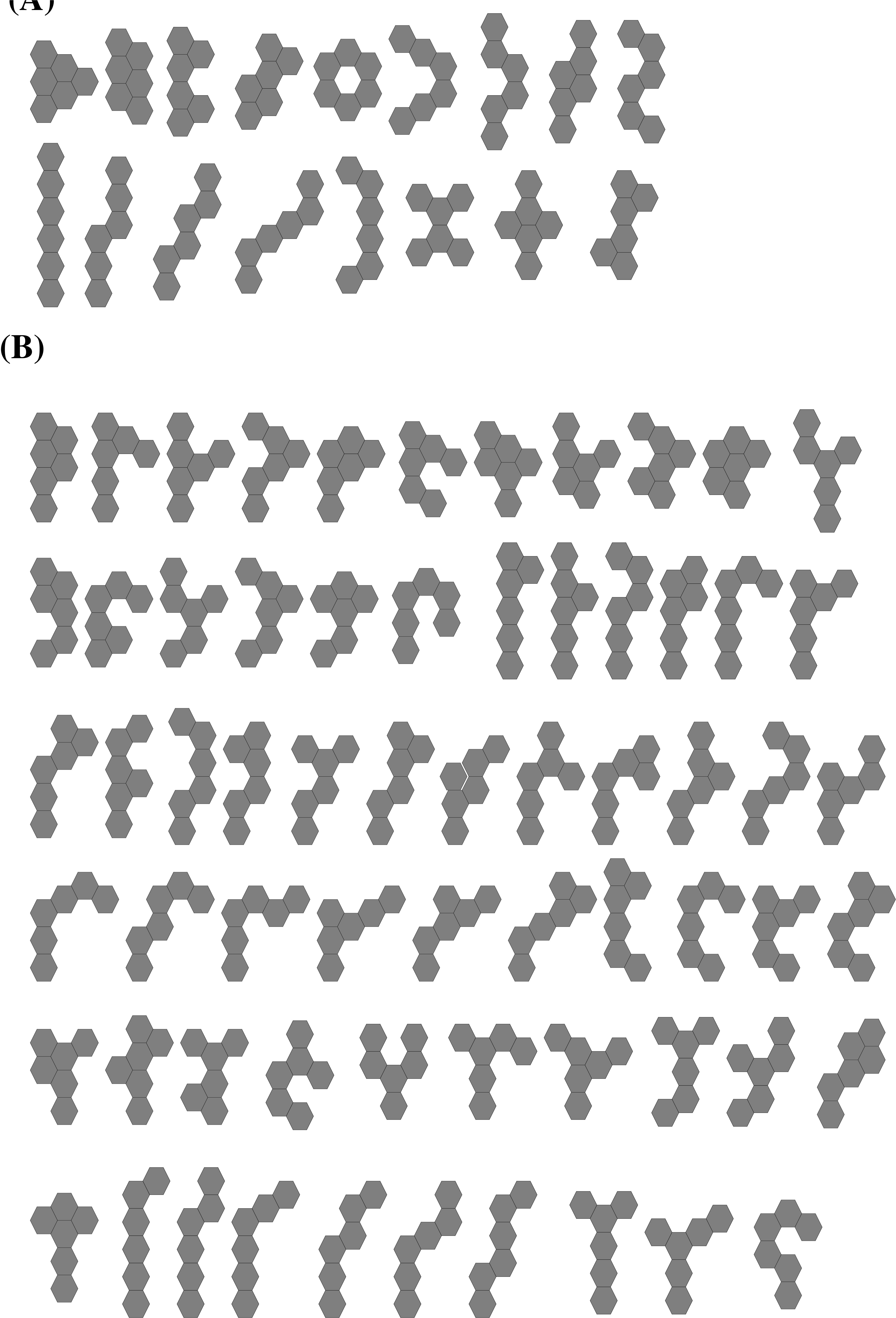}}
\makebox[1cm]{}
\caption{All $82$ lattice animals of size $6$ on a triangular
lattice after symmetry reduction including translational and
$12$ point-group symmetry operations (see text for details).
The centres of the hexagons mark the relative position of the sites of
the triangular lattice on which the spins are flipped with
respect to the model state. See text for a discussion of group A and
group B diagrams.}
\label{fig5}
\end{figure}

\newpage
\begin{figure}
\epsfxsize=11cm
\centerline{\epsffile{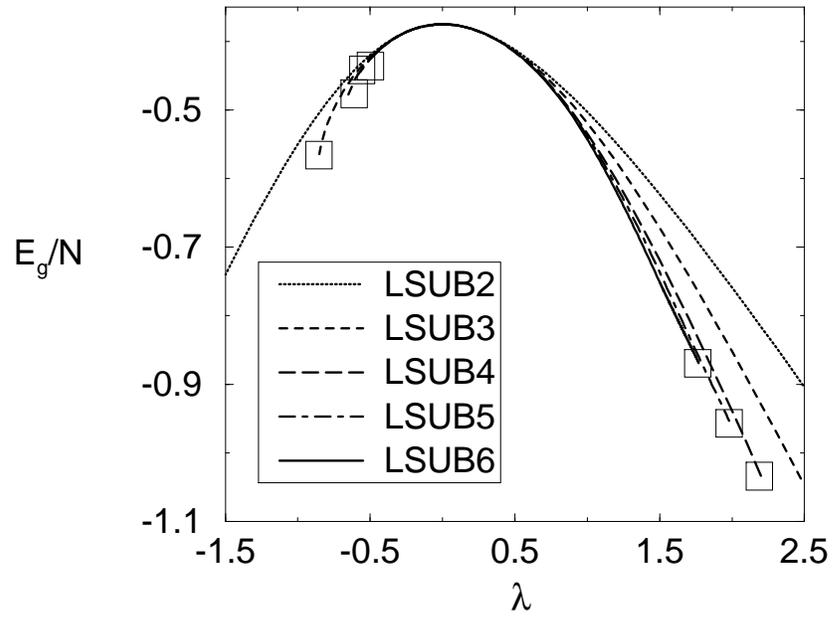}}
\makebox[1cm]{}
\caption{Results for the CCM ground-state energy for the triangular 
HAF. \lsubm critical points $\lambda_{c_1}$ and $\lambda_{c_2}$ 
are indicated by the boxes.}
\label{fig6}
\end{figure}

\newpage
\begin{figure}
\epsfxsize=11cm
\centerline{\epsffile{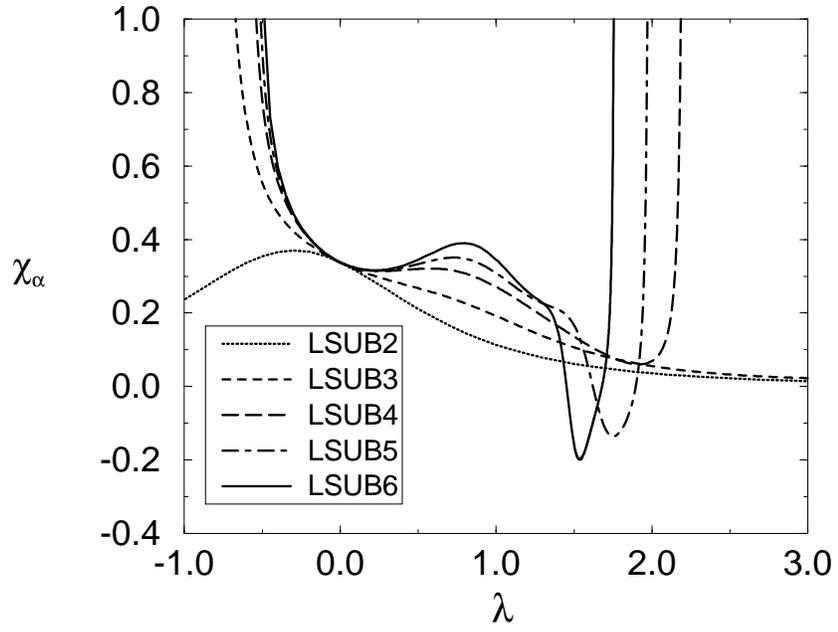}}
\makebox[1cm]{}
\caption{Results for the CCM ground-state anisotropy susceptibility
for the triangular HAF.}
\label{fig7}
\end{figure}

\newpage
\begin{figure}
\epsfxsize=11cm
\centerline{\epsffile{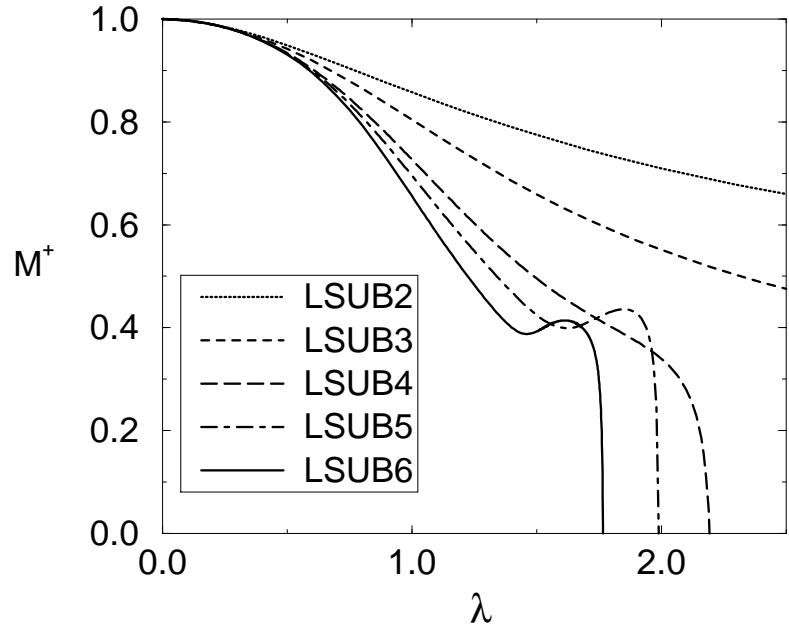}}
\makebox[1cm]{}
\caption{Results for the CCM ground-state sublattice magnetisation
for the triangular HAF.}
\label{fig8}
\end{figure}

\newpage  


\begin{table}
\caption{Results obtained for the spin-$\frac{1}{2}$
\xxz model on the 2D square lattice using CCM \lsubm approximations
($m=2,4,6,8$). $N_{F_1}$ denotes the number of fundamental
configurations for the planar model state, which are further
decomposed in terms of connected and disconnected ones 
respectively, and
$N_{F_2}$ denotes the number of fundamental configurations
for the $z$-axis \neel model state. The ground-state energy per
spin, $E_g/N$, and the sublattice magnetisation, $M^+$, at the
isotropic Heisenberg point ($\Delta=1$) are shown, together with
various critical anisotropy parameters. $\Delta_{c_1}$ and
$\Delta_{c_2}$ indicate the \lsubm critical points for the
planar model state corresponding to the ferromagnetic and
antiferromagnetic phase transitions. $\Delta_{c_3}$ indicates
the critical point for the $z$-axis \neel model state corresponding
to the antiferromagnetic phase transition. Note that there are 
no terminating points in the LSUB$2$ approximation.}
\begin{tabular}{|c|c|c|c|c|c|c|c|}  \hline\hline
$m$     &$N_{F_1}$    	&$N_{F_2}$    	&$E_g/N$ ($\Delta=1$)
        &$M^+$ ($\Delta=1$)             &$\Delta_{c_1}$
        &$\Delta_{c_2}$  &$\Delta_{c_3}$         \\ \hline\hline
2       & 1 (1+0)       & 1 (1+0)       &$-$0.64833
        &0.8414         &  --           & --     & -- \\ \hline
4       & 10 (6+4)      & 7 (5+2)       &$-$0.66366
        &0.7648         &$-$1.249       & 1.648         & 0.577
        \\ \hline
6       & 131 (41+90)   & 75 (29+46)    &$-$0.66700
        &0.7273         &$-$1.083       & 1.286         & 0.7631
        \\ \hline
8       & 2793 (410+2383)& 1287 (259+1028)
        &$-$0.66817
        &0.7048         &  ?           & ?            &0.8429
        \\ \hline
\end{tabular}
\vspace{20pt}
\label{tab:2D_xxz}
\end{table}

\begin{table}  
\caption{Results obtained for the spin-$\frac{1}{2}$
triangular-lattice HAF using CCM LSUB$m$ approximations 
($m=2,3,4,5,6,7$). $N_F$ denotes the number of fundamental
configurations which are further decomposed in terms of connected and 
disconnected ones respectively. Note that only CCM calculations up 
to the LSUB$6$ level of approximation  
are performed in this article. The ground-state energy per spin,  
$E_g/N$, and the sublattice magnetisation, $M^+$, at the 
isotropic Heisenberg point ($\lambda=1$) are shown
for each LSUB$m$ approximation.
The terminating anisotropy parameters, $\lambda_{c_1}$ and 
$\lambda_{c_2}$, which correspond respectively to a phase transition at 
$\lambda$=$-\frac 12$ and another believed [15] to be near $\lambda$=1, 
are also given for each LSUB$m$ approximation. Note that 
there is no terminating point in the LSUB$2$ approximation.}      
\begin{tabular}{|c|c|c|c|c|c|}  \hline\hline 
$m$ 	&$N_F$      	&$E_g/N$ ($\lambda=1$)  &$M^+$ ($\lambda=1$)   
& $\lambda_{c_1}$  & $\lambda_{c_2}$  \\ \hline\hline  
2 	&2 (2+0)     	&$-$0.50290         &0.8578   	
&  --     &  --           \\ \hline
3 	&8 (6+2)  	&$-$0.51911   	 &0.8015   	
&$-$0.86	&5.47     \\ \hline
4 & 30 (16+24)  	&$-$0.53427 	 &0.7273   	
&$-$0.65	&2.20     \\ \hline
5 & 143 (53+90)		&$-$0.53869 	 &0.6958   	
&$-$0.60	&1.98     \\ \hline
6 & 758 (200+558)	&$-$0.54290 	 &0.6561   	
&$-$0.55		&1.77     \\ \hline
7 & 4427 (837+3590)	&?		 &?           	
&?	&?       \\ \hline
\end{tabular}
\vspace{20pt}
\label{t1}
\end{table}

\begin{table}  
\caption{Expansion coefficients in powers of $\lambda$ up 
to the $15$th order for the ground-state 
energy per spin, $E_g/N$, for the anisotropic spin-${1\over 2}$ 
triangular-lattice HAF obtained from the CCM equations 
in the LSUB$6$ approximation. The highest-order known exact 
series expansions up to the $11$th order obtained by 
Singh and Huse [15]
are also included for comparison.}   
\begin{tabular}{|c|c|c|}   
Order 	&LSUB6 			&Exact 		\\ \hline\hline  
0	&$-$0.3750000 		&$-$0.3750000 	\\ \hline
1 	&0.0000000 		&0.0000000 	\\ \hline
2 	&$-$0.1687500 		&$-$0.1687500 	\\ \hline
3 	&0.0337500 		&0.0337500 	\\ \hline
4 	&$-$0.0443371 		&$-$0.0443371 	\\ \hline
5 	&0.0204259 		&0.0204259 	\\ \hline
6 	&$-$0.0283291 		&$-$0.0283291 	\\ \hline
7 	&0.0311703 		&0.0315349 	\\ \hline
8 	&$-$0.0357291 		&$-$0.0476598 	\\ \hline
9 	&0.0541263 		&0.0685087 	\\ \hline
10 	&$-$0.0771681 		&$-$0.1025446 	\\ \hline
11 	&0.1294578 		&0.1565522 	\\ \hline
12 	&$-$0.1848858 		&?   		\\ \hline
13 	&0.2857225 		&?   		\\ \hline
14 	&$-$0.4463496 		&?   		\\ \hline
15 	&0.7021061 		&?   		\\ \hline
\end{tabular} 
\label{t2} 
\end{table}

\end{document}